\documentclass[10pt,a4paper,twocolumn]{scrartcl}
\pdfoutput=1

\usepackage[utf8]{inputenc}
\usepackage[english]{babel}
\usepackage{siunitx}
\usepackage{graphicx}
\usepackage[left=2.15cm, right=2.15cm]{geometry}

\usepackage{booktabs}
\usepackage{caption}
\usepackage[labelformat=empty]{subfig}
\usepackage{amsfonts, amssymb}
\usepackage[nosumlimits]{amsmath}
\usepackage{enumitem}

\usepackage[]{authblk}

\newcommand\shouldeq{\mathrel{\stackrel{\makebox[0pt]{\mbox{\normalfont !}}}{=}}}

\usepackage[numbers]{natbib}
\usepackage{hyperref}
\usepackage{attachfile}

\begin{document}
\title{SNR Spectra as a Quantitative Model for Image Quality in Polychromatic X-Ray Imaging}
\author[a,*]{M. Ullherr}
\author[a,b]{S. Zabler}
\affil[a]{University Würzburg,
Josef-Martin-Weg 63, 97074 Würzburg, Germany}
\affil[a]{Fraunhofer Development Center X-ray Technology EZRT, 
Josef-Martin Weg 63, 97074 Würzburg, Germany}
\affil[*]{Corresponding author: maximilian.ullherr@physik.uni-wuerzburg.de}

\maketitle

\begin{abstract}
In polychromatic x-ray imaging for nondestructive testing, material science or medical applications, image quality is usually a problem of detecting sample structure in noisy data. 
This problem is typically stated this way: As many photons as possible need to be detected to get a good image quality. 

We instead propose to use the concept of signal detection, which is more universal. 
In signal detection, it is the sample properties which are detected. 
Photons play the role of information carriers for the signal. 
Signal detection for example allows modeling the effects which polychromaticity has on image quality.

$\mathit{SNR}$ spectra (= spatial $\mathit{SNR}$) are used as a quantity to describe if reliable signal detection is possible. 
They include modulation transfer and phase contrast in addition to noisiness effects. 
$\mathit{SNR}$ spectra can also be directly measured, which means that theoretical predictions can easily be tested.

We investigate the effects of signal and noise superposition on the $\mathit{SNR}$ spectrum and show how selectively not detecting photons can increase the image quality.
\end{abstract}

\newpage



\section{Introduction}
X-ray imaging light sources are relatively weak light sources, 
so a detection process in x-ray imaging is often noise-limited.
This is especially true for computed tomography (CT), due to the need to acquire many images and the correspondingly longer measurement times.
There are other detection problems in x-ray imaging that are not noise-limited, which we do not consider here. 

It is therefore a natural assumption that, of the photons which pass through the sample, 
the detected fraction needs to be as high as possible to achieve a high image quality. 
Also, there should be as many photons generated as possible.
This can lead to an understanding of image quality where it is implicitly assumed that detecting photons is the main aspect.
As this sets the focus on detecting photons, possibly at the exclusion of other considerations, we will call this concept "photon detection".
Examples for work which (implicitly) use this concept to evaluate image quality are \cite{yaffe1997,cunningham1999,stampanoni2002,Siewerdsen_2002,tanguay2015detective}

While photon detection is a sufficient model in some cases, it is an oversimplification in others:
For example if polychromaticity or modulation transfer are important effects in x-ray imaging. 
Here polychromaticity means that different x-ray energies contribute to a single image.
Generally, if different photons in one image have very different properties, photon detection is an oversimplification. 
To fully model these cases, we fundamentally consider the problem of signal detection in imaging, starting with a problem definition:

\vspace*{1mm}\begin{tabular}{|p{0.89\linewidth}|}
Detection is a measurement process in which a specific signal is reliably differentiated from noise.
\end{tabular}\vspace*{1mm}

To make this definition complete, the terms used are defined and explained in table~\ref{tab:terms}.

\begin{table}[!tp]
\centering
\begin{tabular}{p{0.22\linewidth}p{0.64\linewidth}}
\textbf{term} & \textbf{definition} \\ \toprule
signal & Sample properties and structure which are intended to be measured. \\ \hline
data & Result of a measurement.\\ \hline
noise & Deviation between measured values and signal.\\ \hline
image\quad quality & Degree of reliability with which the sample structure of interest can be detected in a measured image.\\ \hline
modulation  transfer & Image blur (signal deterioration) caused by the detection apparatus (source size, diffraction limit, ...). Quantified by the modulation transfer function (MTF).\\ \hline
poly-chromaticity & Photons have a broad energy distribution.\\
\end{tabular}\vspace*{3mm}
\caption{Definitions for basic terms in the signal detection model.
While it is necessary and useful to define the truth (signal) as a theoretical concept, it is impossible to know exactly. Using photons as information carriers for the signal means that there is always (at least) Poisson noise.
\label{tab:terms}}
\end{table}

The main difference to the concept of photon detection is that we consider the detection of a signal, not the detection of photons. 
Therefore we call this concept "signal detection".
Photons are used to probe the signal, they are information carriers, not the signal itself.

In the case of x-ray imaging, the signal is the spatial distribution (structure) of the x-ray photon interaction strengths of the sample.
The interaction can either be absorption or phase effects, 
which respectively correspond to the imaginary and real parts of the index of refraction.
Usually, only parts of the sample structure is of interest. 
We define the term "image quality" to describe how well these can be detected. 

This work is split into two parts. 
In the first part (section 2), we will take a closer look at what a quantitative model for image quality requires to be sufficiently accurate. 
We then consider if the common approaches satisfy these requirements and explain why SNR spectra are a good quantitative model.
In the second part (the rest of this work), we develop a framework for understanding and evaluating polychromatic image quality based on SNR spectra.

\section{Image Quality Measures}
\begin{figure}[]
\centering
\subfloat[][a) signal example]{\includegraphics[height=0.28\linewidth,angle=-90]{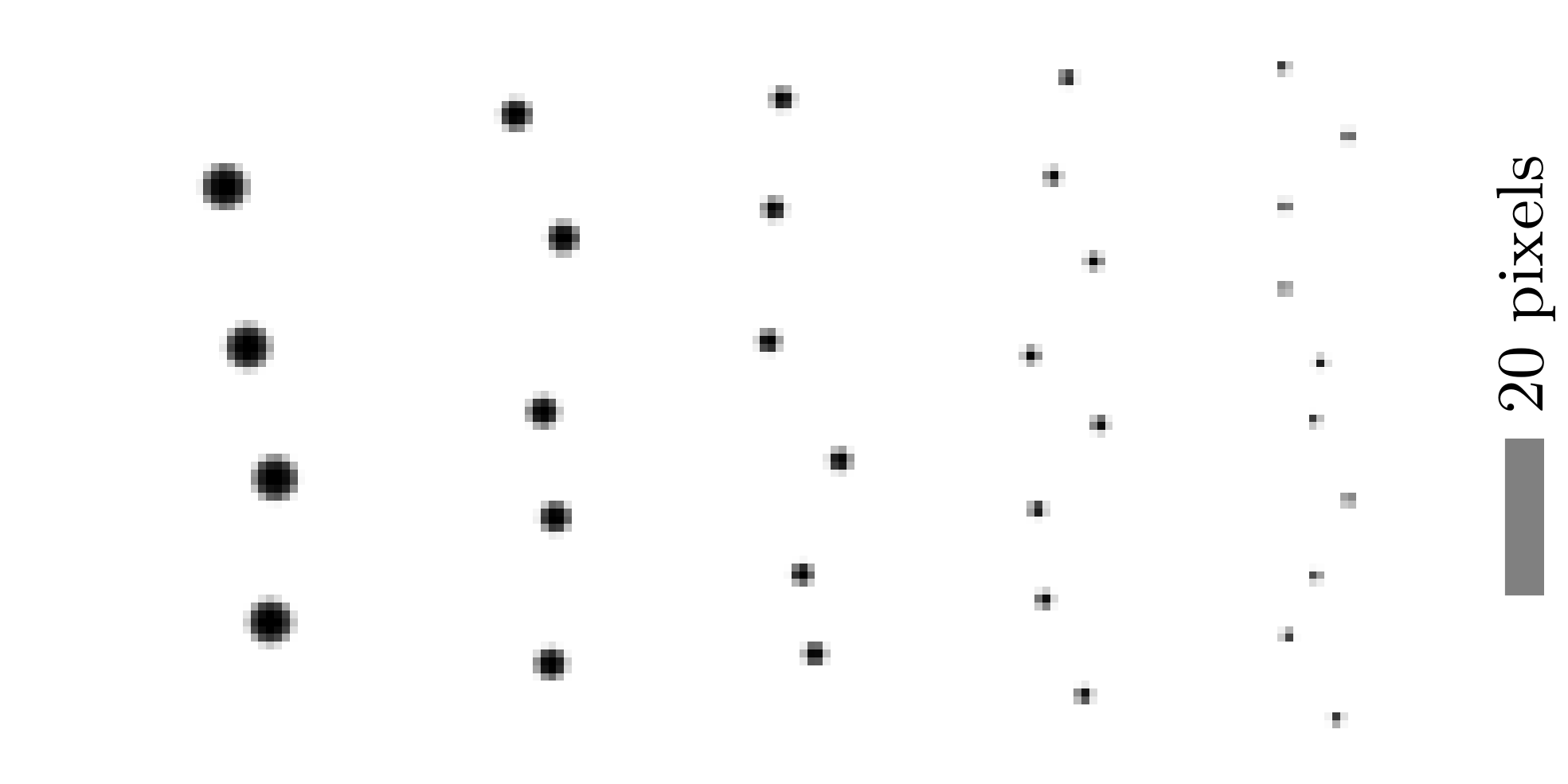}}\ 
\subfloat[][b) superposition\\of signal strengths]{\includegraphics[height=0.28\linewidth,angle=-90]{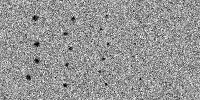}}\ 
\subfloat[][c) superposition\\of different MTFs]{\includegraphics[height=0.28\linewidth,angle=-90]{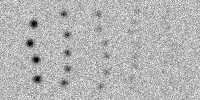}}\\
\subfloat[][d) 1000 counts]{\includegraphics[height=0.28\linewidth,angle=-90]{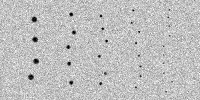}}\ 
\subfloat[][e) 300 counts]{\includegraphics[height=0.28\linewidth,angle=-90]{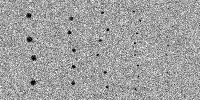}}\ 
\subfloat[][f) 100 counts]{\includegraphics[height=0.28\linewidth,angle=-90]{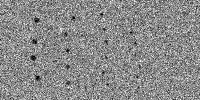}}\\
\subfloat[][g) $\sigma$ = 1.0]{\includegraphics[height=0.28\linewidth,angle=-90]{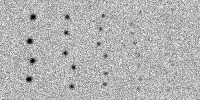}}\ 
\subfloat[][h) $\sigma$ = 2.0]{\includegraphics[height=0.28\linewidth,angle=-90]{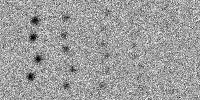}}\ 
\subfloat[][i) $\sigma$ = 4.0]{\includegraphics[height=0.28\linewidth,angle=-90]{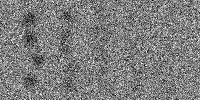}}
\caption{Simulated projection images of randomly placed balls of different sizes with the same number of balls for each size (diameters: 6.0, 4.2, 3.0, 2.1, 1.5).
Images b) and c) are generated by adding another image to d). 
For g)--i), $\sigma$ is the standard deviation of a Gaussian MTF and 1000 counts were used.
\label{fig:detection_effects}}
\end{figure}

\subsection{Effects to Include}
We now want to find a physical model to describe image quality in signal detection quantitatively.
One of the basic difficulties when designing a physical model is deciding which effects to include and which effects to ignore e.g.\ by an approximation. 
A careful consideration of appropriate approximations is required, 
otherwise the physical model will not describe reality correctly in the relevant cases.

The physical model considered in this work is that of a numerical value to describe image quality in a signal detection context. 
As a value to describe a quantity, we will call these different values "measures".\footnote{
Note that these types of measures do not satisfy additivity -- as would be expected from the use in mathematics.} 
These measures are then used to optimize an imaging device or answering the question if one measurement technique is better than another. 
If the measure does not correctly describe the quantity, the wrong case might be considered superior.

\begin{figure}[]
\centering
\includegraphics[width=0.97\linewidth]{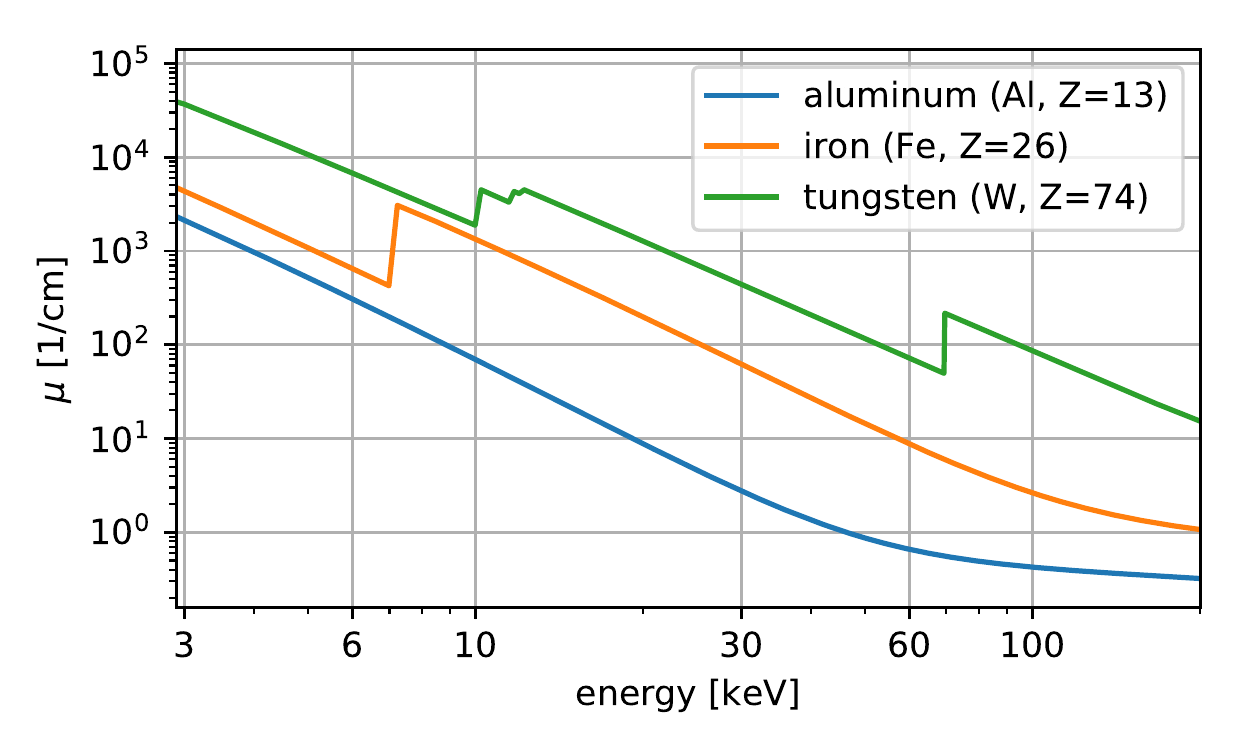}
\caption{Energy dependency of the signal strength for x-ray absorption for some materials. 
A variation by two orders of magnitude is realistic for broad polychromatic spectra. 
Approximating $\mu$ as constant is generally only appropriate for monochromatic or very high energy imaging (see Al curve for $>$ 60\,keV or Fe for $>$ 150\,keV).
This is caused by the fact that Compton scattering becomes the stronger interaction at higher energies and its energy-dependency is weak.
\label{fig:mu_examples}}
\end{figure} 

In polychromatic x-ray imaging, effects from the following categories need to be included:
\begin{itemize}[topsep=4pt,itemsep=3pt,partopsep=3pt, parsep=3pt]
\item[(A)] Noisiness (intensity/Poisson noise and other)
\item[(B)] Modulation transfer (MTF)
\item[(C)] Signal strength (absorption, phase or other)
\item[(D)] Superposition of contributions with different A---C
\item[(E)] Presence of artifacts ($\approx$ noise)
\end{itemize}

Example images for how these effects influence image quality are given in Fig.~\ref{fig:detection_effects}. 
They consist of randomly placed balls of different size but identical maximal absorption length. 
These balls are then blurred with a MTF and Poisson noise is simulated. 
The examples b) and c) are both superpositions of d) and another image with low image quality. 
In both cases, the resulting image has a lower image quality than d) alone.
It can be seen that all effects have the same result: Small objects can or cannot be reliably detected depending on different properties of the imaging setup. 
For example noisiness (A) and MTF (B) have the same resulting effect and should therefore be quantified by a unified measure. 

Note that from Fig.~\ref{fig:detection_effects} d)--f) we can see that small differences in the image quality are not noticeable to a human observer. 
A difference by a factor of three can be differentiated well, much smaller differences would not be. 
This makes image quality optimizations by human estimate very imprecise.

The effects A-C scale the image quality. 
Ignoring one of these effects can give rise to severe misinterpretations. 
Effects D are important if one of the effects A, B or C differs greatly for different detected photons, 
even if the effect themselves would not need to be considered. 
In Fig.~\ref{fig:mu_examples}, an example for a strong energy dependency of the signal strength is shown.
Different intensities (A) and varying signal strengths (C) have the same effect in a single image, but different effects in a superposition.
Otherwise, both effects could be modeled as one effect.
Effects E are included for completeness, but are not considered further in this work.

The two commonly used measures to describe and optimize image quality are $\mathit{DQE}$ (= photon detection efficiency) 
and $\mathit{SNR}$ (= ability/probability to detect signal). 
Both are used in temporal (scalar) and in spatial (frequency-dependent) forms.

\subsection{Signal to Noise Ratio}
The signal to noise ratio gives the strength of the signal in relation to the strength of the noise. 
It can be understood as an (inverse) relative measurement error. 
This concept is based on the fact that signal can only be reliably detected if it is stronger than the noise (relative error $\ll 1$). 
While it is clear that higher $\mathit{SNR}$ is better, defining a sufficiently high $\mathit{SNR}$ depends on many factors and can not be done generally. 
The definitions of the temporal $\mathit{SNR}_t$ and the $\mathit{SNR}$ spectrum are:
\begin{align}
\mathit{SNR}_t &= \frac{\text{pixel\ mean\ signal}}{\text{pixel\ standard\ deviation}} \big(= \mathit{CNR}_t\big) \\
\mathit{SNR}(u) &= \frac{\text{image\ signal\ power\ spectrum}}{\text{image\ noise\ power\ spectrum}} \label{eq:def_SNR_spectrum}
\end{align}

$u$ is the spatial frequency and $\mathit{SNR}(u)$ is the $\mathit{SNR}$ spectrum. 
"Spectrum" here means image spatial frequency spectrum, not light wavelength spectrum.
Note that by "signal", we always mean sample structure, not the light intensity. 
Our definition of $\mathit{SNR}_t$ is identical to a contrast to noise ratio ($\mathit{CNR}_t$) -- the contrast \emph{is} the signal. 
This is consistent with how $\mathit{SNR}(u)$ is defined. 

While $\mathit{SNR}_t$ is suitable to describe the lower noisiness of images with longer integration times, 
comparing cases with differences in the measurement setup (MTF differs e.g.\ due to a different/hardened x-ray spectrum, 
different screen thickness, ...) potentially leads to false comparisons. 
Optimizing $\mathit{SNR}_t$ may thus lead to a lower image quality if this comes at the cost of a worse MTF.

Determining and evaluating both $\mathit{SNR}_t$ and MTF can show this problem.
It does not answer the question as to which case is optimal, which the $\mathit{SNR}(u)$ does.

$\mathit{SNR}_t$ is a linear quantity while $\mathit{SNR}(u)$ is a squared quantity. 
The former is proportional to the signal amplitude while the latter is proportional to measurement integration time.

\subsection{Detective Quantum Efficiency}
A $\mathit{DQE}$ is defined as a $\mathit{SNR}$ transfer function to describe and optimize one part of the imaging device. 
Usually it is applied to describe the x-ray detector. 
It is generally calculated in the form
\begin{align}
\mathit{DQE} = \frac{\mathit{SNR}_\text{out}}{\mathit{SNR}_\text{in}} = \frac{\mathit{SNR}_\text{detected}}{\mathit{SNR}_\text{ideal}}  \in [0, 1[ \label{eq:DQE}
\end{align}

This is intended to give an absolute optimum if $\mathit{DQE} = 1$.
There exist many derived expressions for certain applications (simulation \cite{cunningham1999,Siewerdsen.2004} or measurement\cite{illers2005,IEC-62220-1}). 
The derivations assume monochromaticity but are applied to polychromatic imaging anyway.
They assume that signal strengths cancel out and detector properties can be averaged over the x-ray spectrum. 
The common simplified expression for the $\mathit{DQE}$ of an indirect detector is: 
\begin{equation}
\mathit{DQE}(u) = \frac{a}{1 + c^{-1} H_v(u)^{-2}}\label{eq:def_simple_DQE}
\end{equation}

where $a$, $c$ and $H_v(u)$ are the spectrally averaged x-ray absorption efficiency, x-ray photon conversion factor and optical (scattering) MTF.\footnote{If $a$ is included varies, but it must be included when considering the whole x-ray photon detection process.} 
Effects from different samples are excluded by design in such a simplification, which means that effects (C) and (D) are neglected.

$\mathit{SNR}_\text{ideal}$ and $\mathit{SNR}_\text{detected}$ depend on the sample, which has a x-ray energy dependent transmission. 
This effective x-ray spectrum would need to be considered to correctly compute a polychromatic $\mathit{DQE}$ and can be strongly sample-dependent.

Using a monochromatic $\mathit{DQE}$ at different energies does not pose such problems. 
This is the standard approach for characterizing detectors for optical light (CMOS/CCD) \cite{emva1288}, in which a similar quantity is called quantum efficiency (QE). 
It is the only application of the $\mathit{DQE}$ concept in a polychromatic context which is accurate.

\subsection{Indirect Detection\label{sec:indirect_detection}}
Generally, two types of x-ray imaging detectors can be differentiated:
\begin{itemize}[topsep=4pt,itemsep=3pt,partopsep=3pt, parsep=3pt]
\item Direct detectors: These detectors count single x-ray photon detection events. They are principally able to determine the x-ray energy of a detected photon. They are also called counting detectors.
\item Indirect detectors: These detectors first convert x-ray photons into visible light which is then detected. They usually integrate all events in a certain time frame, weighted by the energy of the x-ray photon. They are also called (energy) integrating detectors.
\end{itemize}

Due to the fact that indirect detectors are currently much cheaper to produce, most x-ray imaging setups use this kind of detector.

For indirect detection, the noise power spectrum $N(u)$ is not a white spectrum, 
but is instead influenced by the detector MTF in the following way \cite{cunningham1994}:
\begin{align}
N(u, E) = I(E) \left[c(E)^2 H_v(u, E)^2 + c(E)\right] \label{eq:indirect_NPS}
\end{align}

where $E$ is the x-ray energy, $H_v(u, E)$ is the optical detector MTF, $I(E)$ is the x-ray intensity and $c(E)$ is the conversion factor for x-rays to visible light.
This equation is only valid for monochromatic x-rays and for a simplified imaging setup, but the general shape of $N(u)$ is identical for polychromatic spectra on real setups. 
In a CT measurement, the reconstruction algorithm will also influence volume image noise.

The pixel noise $\sigma_t$ (temporal standard deviation) for a noise power spectrum $N(u)$ of a projection is given by Parseval's relation \cite{Bracewell_1986}:
\begin{equation}
\sigma_t^2 = \frac{1}{A}\sum_{x, y} N(u_{x, y})\label{eq:temporal_var}
\end{equation}

where $N(u_{x, y})$ is one discrete Fourier coefficient and $A$ is the number of these coefficients.
Computing or measuring a $\sigma_t$ for an indirect detector does not give useful results even for noisiness effects, 
because noise at all frequencies is averaged, although it may not decrease image quality at the structure size that is of interest. 

If the image quality is reduced by decreasing $H_v(u, E)$, 
both $\mathit{SNR}_t$ and $\mathit{DQE}_t$ will increase due to a decreasing $\sigma_t$. 
Changing $H_v(u, E) = 1$ to $H_v(u, E) = 0$ decreases $\sigma_t$ by a factor of $\sqrt{c+1}$. 
If technically feasible, $c > 100$ is achieved in detectors built today. 
This effect can therefore be large. 
In other words, sharper indirect detectors are also temporally noisier. 
The $\mathit{SNR}$ spectrum instead indicates that they have relatively weaker noise, which is correct.

Similar effects are produced by artifacts. 
Beam hardening artifacts or ring artifacts may e.g.\ decrease the $\mathit{SNR}(u)$ at lower frequencies but fine detail can still be detected well. 
A temporal $\mathit{SNR}$ cannot differentiate between low-frequency and high-frequency noise.

\subsection{SNR Spectra}
\begin{table}
\begin{center}\begin{tabular}{l|lllll}
& \multicolumn{5}{c}{effect included}\\
 & A & B & C & D & E \\ \midrule
$\mathit{DQE}_t$ & yes$^*$ & no & no & no & no \\
$\mathit{DQE}(u)$ & yes$^*$ & yes$^*$ & no & no & no \\
$\mathit{SNR}_t$ & yes$^*$ & no & yes & yes$^*$ & no \\
$\mathit{SNR}(u)$ & yes & yes & yes & yes & no
\end{tabular}\end{center}
\caption{Summary of image quality measures and represented physical effects. Entries marked with a $*$ include assumptions or approximations that are incorrect in common cases.\label{tab:measures_effects}}
\end{table}

Which of the image quality measures discussed takes which physical effect into account is summarized in table~\ref{tab:measures_effects}.

Of those measures considered, only $\mathit{SNR}$ spectra are left as a candidate for a generally useful image quality measure -- 
they do not suffer from any of the problems discussed previously. 
While the other image quality measures do work in special cases, 
their restrictions have often been ignored and can lead to severe misinterpretations of physical effects on image quality.

\section{Signal Detection Model}
Understanding image quality in the context of signal detection requires that we model polychromatic effects. 
To do so, we will use $\mathit{SNR}$ spectra. 
These are computed from the fraction of the signal power spectrum to the noise power spectrum, see eq.~(\ref{eq:def_SNR_spectrum}). 
Signal detection image quality can therefore be reduced to the properties of these power spectra, which we will consider in the following.

\subsection{Classes of Photons}
To understand how the different detection properties (e.g.\ due to polychromaticity) affect $\mathit{SNR}$ spectra, we will first consider the signal and noise spectra of photons with the same detection properties. 
This may be e.g.\ a monochromatic thin-screen model which is then used as a building block for the polychromatic thick-screen model. 

Following the use in set theory, we will call this a class of photons: The (detected) x-ray photons that have (approximately) identical detection properties. 
The two most important variables for detection properties are $E$, the x-ray energy and $z$, the position within the scintillation screen. 
The MTF depends on the latter and the depth intensity distribution is energy-dependent.

The actual image then consists of contributions from classes of photons -- the superposition of the images of all individual classes.

\subsection{Signal and Noise Superposition}
To model $\mathit{SNR}$ spectra of polychromatic images, 
the power spectra of signal and noise must be calculated for a superposition of images. 
In the strictest sense, this can be seen as a superposition of x-ray detection events (one image = one event). 
We thus need to derive the superposition equations for signal and noise power spectra.

If $d_\Sigma(x)$ is the polychromatic superimposed intensity image at the pixel position $x$ and $d(x, E)$ the image contribution at energy $E$ (= photon class), we get:
\begin{equation}
d_\Sigma(x) = \sum_E d(x, E)
\end{equation}

where the $\sum$ may also represent an energy integral, in which case $d(x, E)$ is a density. 
In general, this sum may also be a multiple sum/integral over several variables ($E$, $z$, ...), where one combination of variables corresponds to a photon class.

The linear additive noise model can be used to split $d$ into signal $s$ and noise $n$:
\begin{align}
d(x) &= s(x) +  n(x) \label{eq:data_model}
\end{align}

The fact that there is a transformation between physical signal (volume distribution) and $s(x)$ (projection image) is omitted here because it does not influence the results. 
In a volume $\mathit{SNR}$ analysis, the noise is also transformed by the CT reconstruction.
We can derive polychromatic expressions for signal and noise by adding up intensities in image space:
\begin{align}
s_\Sigma(x) &= \sum_E s(x, E)\\
n_\Sigma(x) &= \sum_E n(x, E)
\end{align}

The image power spectrum is calculated as:\footnote{In the following we denote power spectra by capital letters, not the Fourier transformed quantities (except for the MTF $H$). Note that if power spectra are measured, corresponding measurement errors occur.}
\begin{align}
D_\Sigma(u) &= \left| \mathcal{F}\left\{ d_\Sigma(x) \right\}(u)\right|^2 = S_\Sigma(u) + N_\Sigma(u)\label{eq:power_spectrum}
\end{align}

because noise is uncorrelated with any other signal/noise and the corresponding mixed terms vanish. 
Similarly, for the noise power spectrum itself we get:
\begin{align}
N_\Sigma(u) &= N_0(u) + \left|  \mathcal{F}\left\{ \sum_E n(x, E) \right\}(u)\right|^2 \\
&= N_0(u) + \sum_E N(u, E) \label{eq:noise_sum}
\end{align}

where $N_0$ are noise contributions in addition to photon noise, e.g.\ camera readout/dark noise. Scattered photons are counted in the noise sum (energy of the scattered photon), but usually do not contribute to the signal.

For the signal, we get:
\begin{align}
S_\Sigma(u) &= \left|  \mathcal{F}\left\{ \sum_E s(x, E) \right\}(u)\right|^2,\label{eq:signal_sum_base}
\end{align}
which is simplified in the next section.

\subsection{Local Area Approximation}
If we consider the image quality in a small area (locally), we can approximate the intensity as constant within that area. 
In the following, intensity thus stands for the number of x-ray photons (units of 1) detected in a local area. 
A local area is much larger than a single pixel and contains weak sample structure. 
This also allows us to assume that for absorption, the signal strength  is proportional to the absorption strength. 
The proportionality is given by a linear approximation of the absorption curve (Lambert-Beer law).

Because in the end we are interested in finding an optimum of the unit-free quantity $\mathit{SNR}$, physical units are not considered. 
They can be restored if needed. 
We can thus write a simplified model for the signal and its power spectrum as: 
\begin{align}
s(x, E) &=  I(E) \left[\kappa(E) \rho(x)\right] \,*\, h(x, E)\\
S(u, E) &= P(u)\ \big[ I(E)  \kappa(E) H(u, E) \big]^2
\end{align}

The signal is a product of the detected intensity $I(E)$, an energy independent spatial matter density $\rho(x) \leftrightarrow P(u)$, the MTF $H(u, E)$ and an interaction strength $\kappa(x, E)$. 
We have used $\kappa(x, E) = \kappa(E)$ (single material sample) for simplicity---considering multi material samples would not give new effects. 
Note that we assume the imaging device to be a linear translationally invariant (LTI) system in the local area.

$\kappa(E)$ can model almost any signal generating physical effect. 
This is usually x-ray absorption or phase contrast. 
In an cases like imaging with x-ray optics or grating interferometry \cite{paganin2006}, $\kappa(x, E)$ is also determined by the energy band for which the x-ray optics or gratings are optimized.

For a signal sum eq.~(\ref{eq:signal_sum_base}) becomes:
\begin{align}
S_\Sigma(u) &= \mathrm{P}(u)  \left[\sum_E I(E) \kappa(E) H(u, E)\right]^2 \label{eq:signal_sum}
\end{align}
The influence of the sample structure $P(u)$ is a global scaling factor. 
The right hand side originally contains the square of the absolute value, but all sum contributions are real valued.
It can be easily seen that signal sums are difficult to evaluate, 
because many physical effects need to be considered---even in a basic model.

Signal power spectra of large structures show higher amplitude at lower  spatial frequencies and $P(u)$ is then usually monotonically decreasing towards higher frequencies. 
The reason is that sample structures are long-range correlations in the signal, and the Wiener-Chinchin-Theorem translates this to the power spectrum shape.
Larger structural details therefore have an intrinsically higher $\mathit{SNR}$ spectrum value and are easier to detect.
A lower $\mathit{SNR}(u)$ usually has the effect that the detail resolution of a measurement gets worse.

It is possible to define a quantity that is not dependent on the amount of structure in the sample or on the shape of the signal power spectrum. 
We will call this the detection effectiveness spectrum, which is derived from eq.~(\ref{eq:noise_sum}) and eq.~(\ref{eq:signal_sum}) and defined as:
\begin{align}
\mathit{DE}(u) &= \frac{\mathit{SNR}_\Sigma(u)}{P(u) \eta} =\frac{S_\Sigma(u)}{N_\Sigma(u) P(u) \eta} \nonumber \\
&= \frac{ \big[\sum_E I(E) \kappa(E) H(u, E)\big]^2 }{\big[N_0(u) + \sum_E N(u, E)\big]\eta} \label{eq:DE}
\end{align}

Introducing the cost function $\eta$ further allows to compare measurements with e.g.\ different measurement times ($\eta$ = time). 
In medical imaging, using the dose as a cost function can be appropriate. 
Optimizing the $\mathit{SNR}(u)$ will give the same result as optimizing the $\mathit{DE}(u)$ if $\eta$ is constant. 
In the following considerations, the $\mathit{SNR}(u)$ could usually be replaced by the $\mathit{DE}(u)$.

The $\mathit{DE}(u)$ does not depend on the specific sample structure and thus describes only the imaging setup itself. 
Its main advantage compared to the $\mathit{SNR}(u)$ is that the $\mathit{DE}(u)$ allows comparable measurements. 
In contrast to e.g.\ a $\mathit{DQE}$, this quantity models polychromatic effects.
Sample-dependent properties of an imaging setup are included by the $\mathit{DE}(u)$.

\subsection{SNR Spectra of Superpositions\label{sec:SNR_sums}} 
One important aspect of $\mathit{SNR}$ spectra are the effects that come from superimposing images with different properties, e.g. different $\kappa$ or MTF. 
From eq.~(\ref{eq:noise_sum}) and eq.~(\ref{eq:signal_sum}) we get:
\begin{equation}
\mathit{SNR}_\Sigma(u) =  P(u)\frac{   \big[\sum_k I_k \kappa_k H_k(u)\big]^2 }{N_0(u) + \sum_k N_k(u)} \label{eq:SNR_superposition}
\end{equation}

where $k$ is the index for different contributions to the resulting image -- 
a contribution may consist of a single photon class or can itself be a sum of photon classes.
While both sums are linear superpositions, 
the square in the numerator makes many simplifications impossible. 
An important property of eq.~(\ref{eq:SNR_superposition}) is that the signal contributions are weighted with the signal strength $\kappa$ and MTF $H$, while the noise contributions are not. 

We can see from eq.~(\ref{eq:SNR_superposition}) that adding two images with greatly differing $\mathit{SNR}$ (due to relative differences in signal strength or MTF) will result in an image with a lower $\mathit{SNR}$ than that of the one image with the higher $\mathit{SNR}$.
The following triangle inequality holds:
\begin{equation}
\mathit{SNR}_{1+2}(u) \leq \mathit{SNR}_1(u) + \mathit{SNR}_2(u)
\end{equation}

Detecting more photons (image 2) in addition to the reference situation (image 1) increases the overall $\mathit{SNR}$ if:
\begin{align}
\mathit{SNR}_{1+2}(u) &> \mathit{SNR}_1(u) \\
\Leftrightarrow \frac{ \big[ I_1 \kappa_1 H_1 + I_2 \kappa_2 H_2\big]^2 }{N_0 + N_1 + N_2} &> \frac{ \big[ I_1 \kappa_1 H_1\big]^2 }{N_0 + N_1}\\
\Leftrightarrow \frac{I_1 \kappa_1 H_1 + I_2 \kappa_2 H_2}{I_1 \kappa_1 H_1} &> \sqrt{\frac{N_0 + N_1 + N_2}{N_0+N_1}}\\
\Leftrightarrow \frac{I_2 \kappa_2 H_2}{I_1 \kappa_1 H_1} &> \sqrt{1 + \frac{N_2}{N_0+N_1}} - 1 \label{eq:SNR_gain_condition}
\end{align}

If this condition is violated, the additional photons (image 2) would decrease $\mathit{SNR}$ if they are detected. 
In this equation, the absolute signal strengths are not important, but the relative strengths are.

For an infinitely small addition ($N_2 \ll N_1$) we can approximate the square root above and get:
\begin{align}
\frac{I_2 \kappa_2 H_2}{N_2} &> \frac{1}{2}\frac{I_1 \kappa_1 H_1}{N_1 + N_0} \label{eq:SDS_I_0}
\end{align}

While this is an interesting limit case, it does not have any practical application.
We define these naturally arising quantities as the "signal detection strength":
\begin{align}
\mathit{SDS}(u, E) &= \frac{I(E) \kappa(E) H(u, E)}{N(u, E)} \label{eq:SDS}
\end{align}

A $\mathit{SDS}$ can be calculated either for a photon class or for a set of photon classes ${\{E\}}$: 
\begin{equation}
\mathit{SDS}(u, \{E\}) = \frac{\sum_{\{E\}} I(E) \kappa(E) H(u, E)}{\sum_{\{E\}} N(u, E)} \label{eq:class_SDS}
\end{equation}

For $N_0 \ll N_1$ we can rewrite eq.~(\ref{eq:SDS_I_0}) to:
\begin{equation}
\mathit{SDS}_2(u, E) > \frac{1}{2} \mathit{SDS}_1(u, E) \label{eq:SDS_half_condition}
\end{equation}

For an ideal detector ($I = N$), the $\mathit{SDS}$ simplifies to:
\begin{equation}
\mathit{SDS}(u, E) = \kappa(E) H(u, E)
\end{equation}

Note that for absorption imaging, $H(u, E) \leq 1$ and $\mathit{SDS}(u, E) \leq \kappa(E)$.
For phase contrast imaging, $H(u, E)$ may be $> 1$ (or even $\gg 1$).

From eq.~(\ref{eq:SDS_half_condition}) we can see that a small intensity addition requires at least half the $\mathit{SDS}$ of the existing image to contribute positively. 
On the other hand, for a large intensity addition ($I_2 \gg I_1$), $\mathit{SNR}_{1+2} > \mathit{SNR}_1$ is always true. 
As an example for an intermediate case, if $N$, $H$ and $I$ are equal for both images, 
the condition becomes $\kappa_2 > 0.41 \kappa_1$.

\subsection{Weighted Superposition\label{sec:energy_weighting}}
As discussed in \cite{Tapiovaara_1985}, applying an energy-dependent weighting factor before the superposition of the image contributions can increase the $\mathit{SNR}$. 
This is called "energy weighting" and there are two variants:

\begin{itemize}[topsep=4pt,itemsep=3pt,partopsep=3pt, parsep=3pt]
\item Multiplication with an energy-dependent weighting factor after the detection of the x-ray photons. 
This is what is usually discussed as energy weighting. 
Because sufficiently good energy-resolving 2D imaging detectors are currently not technically feasible, this is mostly a theoretical concept. \\
We will call this effect "computational energy weighting" (CEW).\\
A CEW weight $w$ scales $\mathit{SDS}$ with $1/w$.
\item Implicit weighting by the different detected intensities $I(E)$ (effective x-ray spectrum). 
This is influenced e.g.\ by the source spectrum or the detector absorption efficiency spectrum.
This weighting can be changed by purposefully decreasing the detected intensity e.g.\ in specific energy ranges.\\
We will call this effect "detection energy weighting" (DEW). 
\end{itemize}

In both cases, optimizing the weighting can lead to an increase in $\mathit{SNR}$. 
The increases vary for different experimental conditions between a few percent and factors~$> 10$. 
The larger effect sizes are for very broad spectra that are not useful without energy weighting.

For the CEW, if we multiply with a weight $w(u, E)$, we get:
\begin{equation}
\mathit{SNR}_\text{CEW}(u, w) = \mathrm{P(u)} \frac{ \big[\sum w(u, E) I(E) \kappa(E) H(u, E)\big]^2 }{\sum w(u, E)^2 N(u, E)}
\end{equation}

To compute the optimal energy weighting function $w_\text{opt.}$ for CEW, 
we determine the maximum of the $\mathit{SNR}$ spectrum depending on $w(u, E)$:
\begin{equation}
\frac{\mathrm{d}\mathit{SNR}_\text{CEW}(u, w)}{\mathrm{d}w} \shouldeq 0
\end{equation}

which is solved by any CEW weight proportional to
\begin{equation}
w_\text{opt.}(u, E) \propto \frac{I(E) \kappa(E) H(u, E)}{N(u, E)} =  \mathit{SDS}(u, E) \label{eq:CEW_optimal_weight}
\end{equation}

This is an extension of the known result $w_\text{opt.}(u, E) \propto \kappa(E)$ \cite{Tapiovaara_1985}. 
For a detector with a limited energy resolution, eq.~(\ref{eq:class_SDS}) can be used.
To avoid artificial image blurring, 
it is sufficient to only apply a relative blurring factor in this way:
\begin{equation}
w_\text{opt.}(u, E) = \frac{I(E) \kappa(E) H(u, E)}{N(u, E)H_0(u)} 
\end{equation}

where $H_0(u)$ can e.g.\ be the energy-averaged MTF.
Note that the $\mathit{SNR}$ spectrum is not changed when a Fourier filter is applied, but the image may be easier or more difficult to interpret. 
After application of the optimal CEW, the signal detection strength $\mathit{SDS}$ is constant with respect to $E$:
\begin{equation}
\mathit{SDS}_\text{opt.\ CEW}(u, E) = \text{const}(E) \label{eq:CEW_SDS}
\end{equation}

The special case of a task-independent ($\text{t.i.}$) measurement case is defined as a case with energy independent signal detection strength (without weighting):
\begin{align}
\mathit{SDS}_\text{t.i.}(u, E) &= \text{const}(E) \label{eq:task-indepenedent_SDS} \\
\Rightarrow w_{\text{opt}|\text{t.i.}}(u, E) &= \text{const}(E) \label{eq:task-independency}\\
\Rightarrow \mathit{SNR}_{\Sigma|\text{t.i.}}(u) &= \sum_E \mathit{SNR}_\text{t.i.}(u, E)
\end{align}

In this case, the optimal $\mathit{DQE}$ also gives the optimal $\mathit{SNR}$. 
This is the simplest possible case which is often implicitly used to understand detection efficiency. 
To do so, eq.~(\ref{eq:task-indepenedent_SDS}) must be assumed to hold true approximately.
This is only appropriate in some special cases, e.g.\ monochromatic imaging.

If we compare eq.~(\ref{eq:CEW_SDS}) and eq.~(\ref{eq:task-indepenedent_SDS}), we can see that applying an optimal CEW has the effect that the result of the CEW appears to be task-independent.
The CEW itself is of course task-dependent.
This also means that task-independency is the case in which there is an optimal superposition of the $\mathit{SNR}$ contributions.

For the optimal DEW, there appears to be no analytical expression.
The optimal DEW is thus the case with the maximal $\mathit{SNR}$.
Eq.~(\ref{eq:SNR_gain_condition}) can be used to find approaches that may increase $\mathit{SNR}$.
The general principle is that if the x-ray photons at specific energy ranges do not contribute positively to $\mathit{SNR}$, a reduction of their intensity increases $\mathit{SNR}$.

\section{Examples}
\subsection{Two Images Superposition\label{sec:sde_sum_example}}
\begin{table}
\centering
\begin{tabular}{llll}
 description & $\kappa_2$ & $\mathit{SNR}_2$ & $\mathit{SNR}_{1+2}$ \\ \toprule
a) photoabsorption & 0.14 & 0.020 & 0.65 \\
b) phase contrast & 0.25 & 0.063 & 0.78 \\
c) no gain, eq.~(\ref{eq:SNR_gain_condition}) & 0.41 & 0.17 & 1 \\
d) task-independent & 1 & 1 &2 \\
e) optimal CEW for a) & 0.14 & 1 & 1.02 \\
f) optimal DEW for a) & 0.14 & 1 & 1
\end{tabular}
\caption{Examples for the $\mathit{SNR}(u_a)$ of image sums for different physical effects represented by $\kappa$ with $\kappa_1$ = 1 and $\mathit{SNR}_1$ = 1.
Cases a) and b) are e.g.\ iron for 30\,keV and 60\,keV. 
Note that for a) to d), the fraction of detected photons ($\propto$~polychromatic $\mathit{DQE}$) is twice as high for the sum image, but for a) and b) the image quality is lower. 
See section~\ref{sec:energy_weighting} for cases e)$+$f).\label{tab:SDE_additivity}}
\end{table}

\begin{figure}[]
\centering
\includegraphics[width=0.97\linewidth]{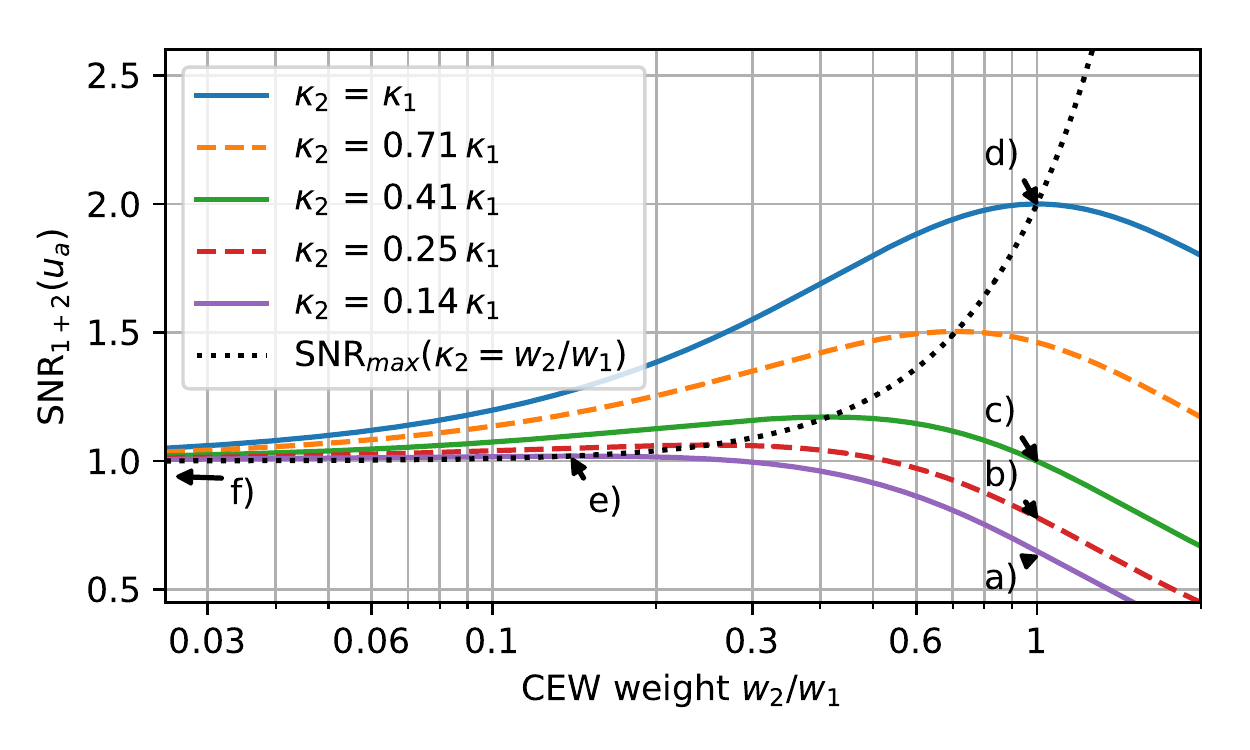}
\includegraphics[width=0.97\linewidth]{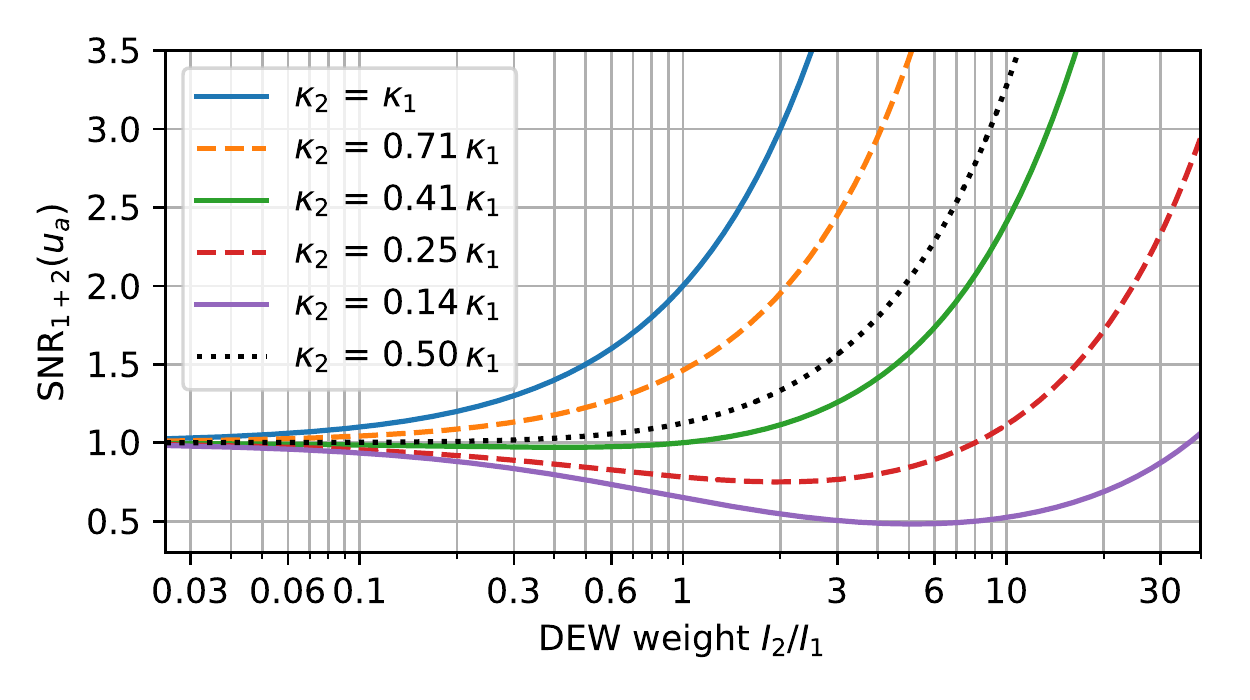}
\caption{Examples for the dependency of the $\mathit{SNR}$ on the weighting factor for CEW (top) and DEW (bottom). 
The weighting factor factor of image 2 is $w_2$. 
The annotations correspond to table~\ref{tab:SDE_additivity}. 
For DEW, the maximal $\mathit{SNR}$ is either at $I_2 = 0$ or at finite $I_2$, depending on the maximal value of $I_2$.
The edge case $\kappa_2 = 0.5\kappa_1$ is the lowest possible $\kappa_2$ where $\mathit{SNR}_{1+2} \geq \mathit{SNR}_1$ for all $I_2$.
\label{fig:SNR_for_CEW_DEW}}
\end{figure}

As the simplest possible example for $\mathit{SNR}$ spectra of an image superposition, we will consider adding two monochromatic images:
\begin{equation}
d_{1+2}(E) = d_1 \delta(E-E_1) + d_2 \delta(E-E_2)
\end{equation}

We assume the following conditions:
\begin{align}
&P(u_a) H_{1}(u_a)^2 = P(u_a) H_{2}(u_a)^2 = 1;\\
&I_1 = I_2 = 1;\ \ E_1/E_2 = 0.5
\end{align}
for a structure size of $u_a$. 
This effectively excludes modulation transfer effects from our examples---this can be done in a theoretic model without loss of generality,\footnote{
Including $H(u_a)$ as a factor in $\kappa$ restores the full properties.}
 but would be almost useless for a real application.

This is a realistically polychromatic case whose effect sizes approximate real cases. 
Inserting all these values into equations~(\ref{eq:noise_sum}) and (\ref{eq:signal_sum}) yields:
\begin{align}
N_{1+2}(u_a) &= I_1 + I_2 \\
S_{1+2}(u_a) &= (I_1 \kappa_1 + I_2 \kappa_2)^2\\
N_{1}(u_a) &= I_1; &N_{2}(u_a) &= I_2 \\
S_{1}(u_a) &= (I_1 \kappa_1)^2; &S_{2}(u_a) &= (I_2 \kappa_2)^2
\end{align}

where we assume an ideal photon counting detector. 
Using the specific values listed above gives:
\begin{align}
\mathit{SNR}_1 &= \mathit{SNR}_1(u_a) = 1 \\
\mathit{SNR}_2 &= \kappa_2^{\ 2} \\
\mathit{SNR}_{1+2} &= \frac{(1 + \kappa_2)^2}{2}
\end{align}

For some examples of $\kappa_2$ the $\mathit{SNR}$ values for image sums are computed in table~\ref{tab:SDE_additivity}. 
From these examples we can see that the additionally detected intensity $I_2$ may reduce the overall image quality significantly if the $\mathit{SNR}_2$ of this additional intensity is low compared to the $\mathit{SNR}_1$ of $I_1$. 
A difference in a MTF may also produce this effect.

The reason for this behavior is easily explained by the fact that 
if $\kappa_1^2 \gg \kappa_2^2$, $I_2$ contributes little signal but all of its noise to the image sum. 
A lower detected intensity at $E_2$ would decrease the noise.

For this example, a high enough intensity can compensate for a low signal strength in the following way:
\begin{equation}
\mathit{SNR}_{1+2} \geq \mathit{SNR}_1 \Leftrightarrow I_2 \geq I_1 \frac{1 - 2\kappa_2}{\kappa_2^{\ 2}} \label{eq:DEW_intensity_gain_point}
\end{equation}

This simple example allows us to evaluate the effect (non-optimal) energy weighting has on the $\mathit{SNR}$,
which is shown in Fig.~\ref{fig:SNR_for_CEW_DEW} for CEW and DEW. 
The examples only qualitatively represent realistic cases, but demonstrate the effects which different energy weightings have on $\mathit{SNR}$. 

We can see that for CEW, there is always a unique maximum at $w_2 = \kappa_2$. 
For DEW, there is a unique minimum, while the maximum is obtained for $I_2 \rightarrow \infty$, and for $\kappa_2 < \kappa_1/2$ an additional local maximum exists at $I_2 = 0$. 
If we assume that a maximal value for $I_2$ is given by the physical circumstances (e.g.\ source spectrum),  
the optimal weighting is either to use this maximal value or use $I_2 = 0$ (depending on $\kappa_2$). 
The optimal DEW weight can thus be interpreted as a mask function that is either 0 or 1. 
This is usually a step function that is 1 at low energies and 0 above some threshold.

\subsection{Simple Polychromatic Simulations\label{sec:sps}}
\begin{figure}[]
\centering
\includegraphics[width=0.97\linewidth]{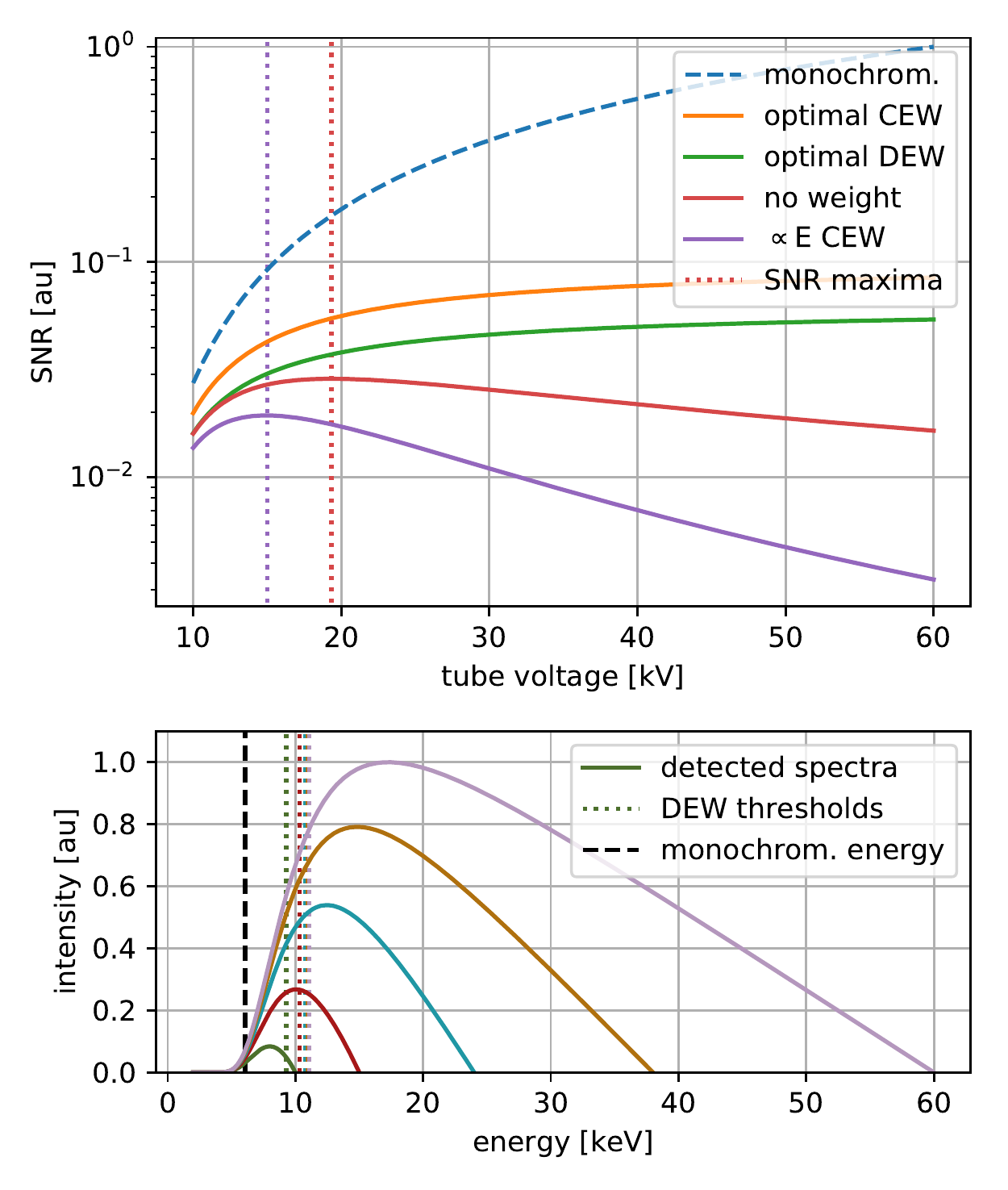}
\includegraphics[width=0.97\linewidth]{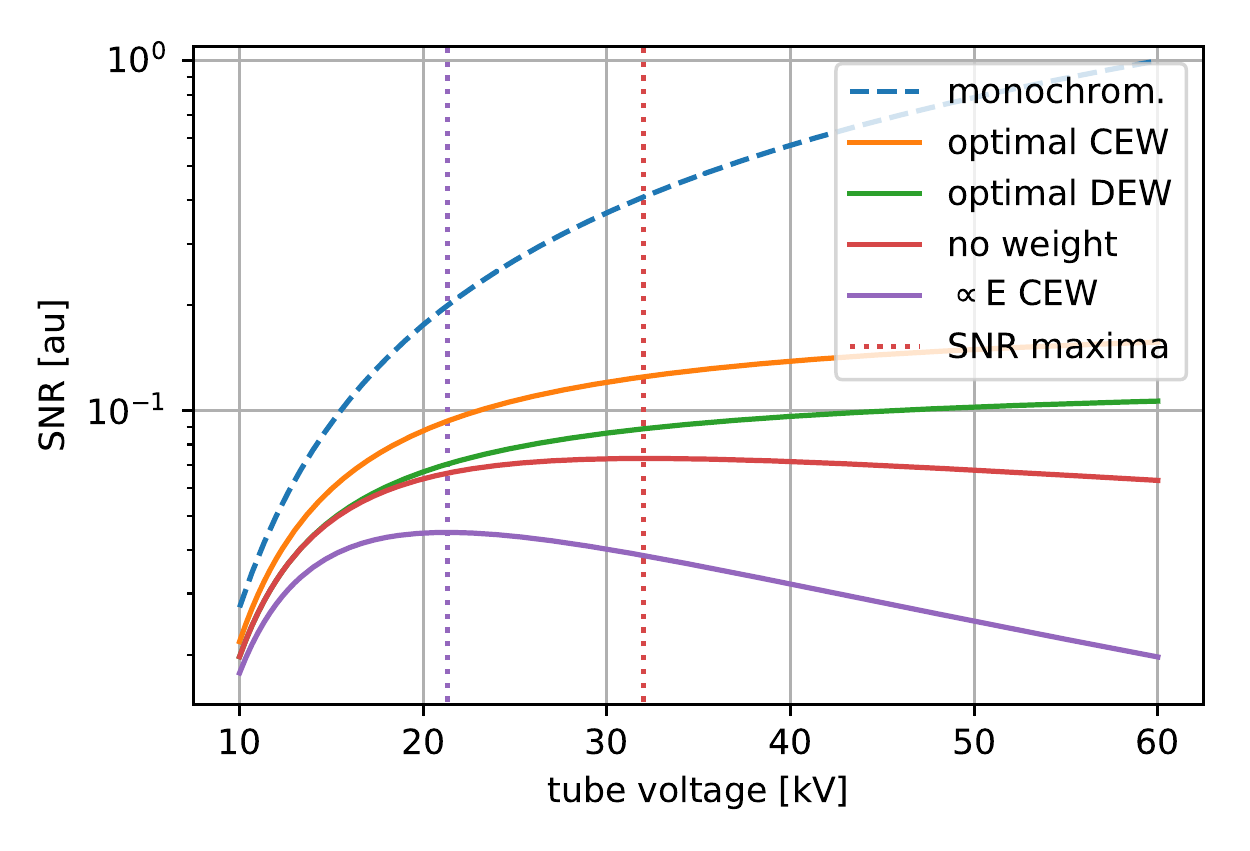}
\caption{Simulated (photo-)absorption $\mathit{SNR}(u_a)$ (top) and x-ray inline phase contrast $\mathit{SNR}(u_a)$ (bottom) curves for a bremsstrahlung source with constant source power. 
The corresponding detected x-ray spectra are shown in the middle, where the highest energy in a curve is the tube voltage. 
\label{fig:SNR_sim_example}}
\end{figure}

$\mathit{SNR}$ simulations for simple x-ray spectra are given in Fig.~\ref{fig:SNR_sim_example}. 
They are generated by analytically computing $\mathit{SNR}(u)$, using eq.~(\ref{eq:SNR_superposition}).
Both examples use the same source spectra, detector and sample, but different energy-dependencies of the signal strength. 
These examples demonstrate the magnitude of the effect different energy weighting effects have on $\mathit{SNR}$, they are not meant to describe an actual setup.
See Fig.~\ref{fig:mu_examples} for the absorption coefficient for aluminium (Al), which gives the energy-dependent sample transparency (\SI{0.1}{\milli\meter} thickness) and the signal strength for the absorption signal (approximately $\kappa(E) \propto E^{-3}$). 
The x-ray inline phase contrast signal strength is assumed to be $\kappa(E) \propto E^{-2}$ \cite{paganin2006}. 

An ideally absorbing detector is used and the source is assumed to have constant target power (product of tube voltage and current is constant). 
Its spectrum is a bremsstrahlung spectrum given by Kramers' law, which is sufficient as a rough approximation. 
Higher tube voltages correspond to a higher degree of polychromaticity.

To produce plots that are easy to interpret, we again evaluate the $\mathit{SNR}$ spectrum at a specific spatial frequency $u_a$ and assume an energy-independent MTF. 
Applications in real world examples must avoid this kind of simplification. 

The monochromatic $\mathit{SNR}$ curve is from the optimal monochromatic x-ray spectrum (at 6.1\,keV for absorption and 7.0\,keV for phase contrast) with the intensity given by the sum over the corresponding bremsstrahlung spectra.\footnote{
To get realistic monochromatic intensities, $\frac{1}{3}$ of the sample thickness is counted as a x-ray filter.} 
The optimal CEW curve uses the signal strength as the weight. 
The optimal DEW curve only includes photons below the optimal threshold, shown as vertical dashed lines in Fig.~\ref{fig:SNR_sim_example} (middle). 
The "$\propto E$ CEW" is the energy integrating indirect detector case ($w \propto E$) and assumes efficient conversion (see eq.~(\ref{eq:indirect_NPS}), $c^2H_v^2 \gg c$). 

Both for photoabsorption and phase contrast signals, only the lower energy x-ray photons have a sufficiently high $\mathit{SDS}$ to contribute positively to the $\mathit{SNR}$. 
This effect is weaker for phase contrast due to a weaker energy dependency.
In this example, the $\mathit{SNR}$ curves are always ordered: $\propto E$ CEW $<$ no weight $\leq$ optimal DEW $<$ optimal CEW $<$ monochromatic.
The differences get larger for broader x-ray spectra.

Direct detectors ("no weight") here have an intrinsically higher image quality than indirect detectors ("$\propto E$ CEW") and this benefit is larger for higher degrees of polychromaticity. 
This difference is caused solely by the different energy weighting, as both detectors are otherwise assumed to be perfect absorbers and without additional noise. 

\begin{figure}[]
\centering
\includegraphics[width=0.97\linewidth]{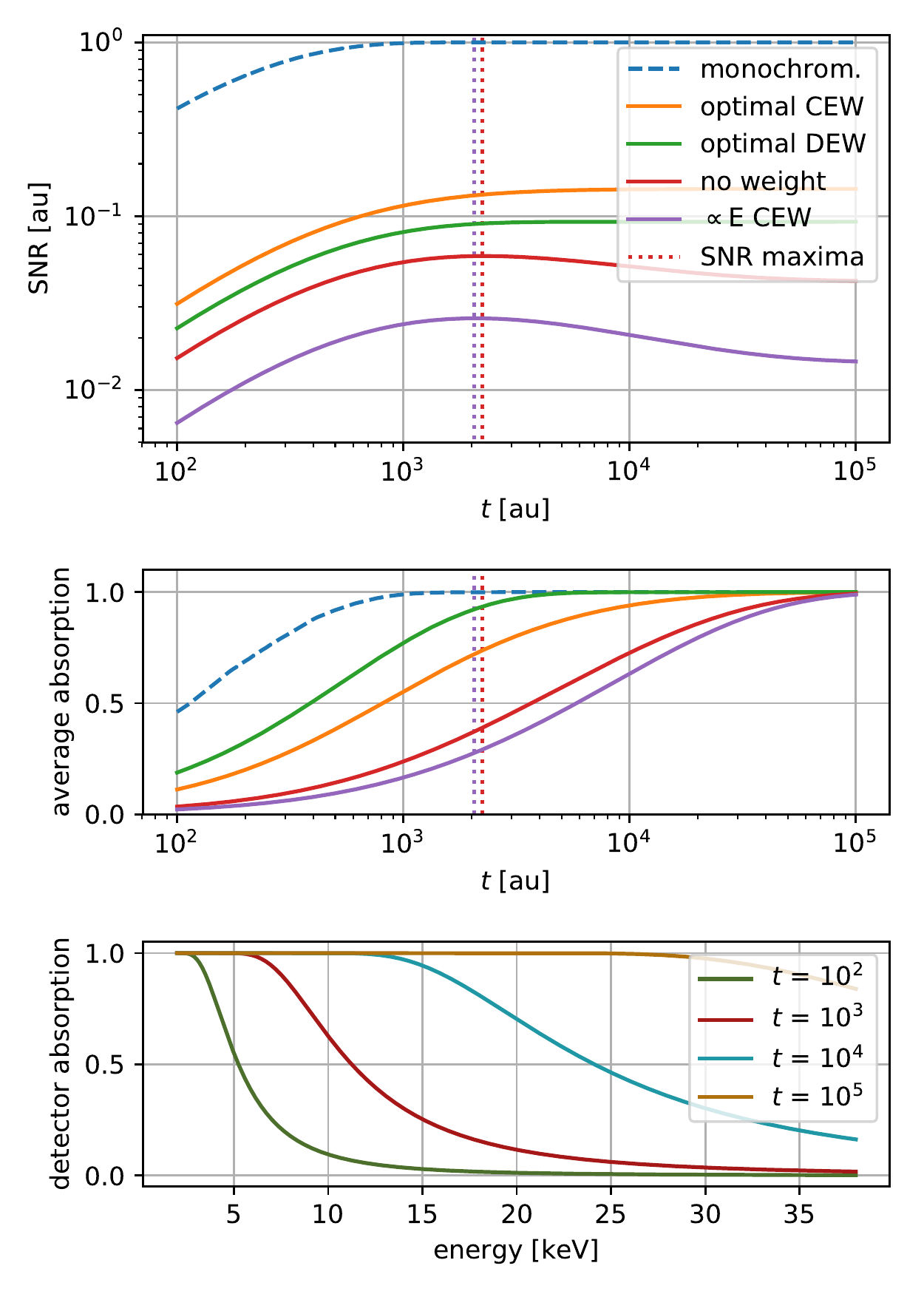}
\includegraphics[width=0.97\linewidth]{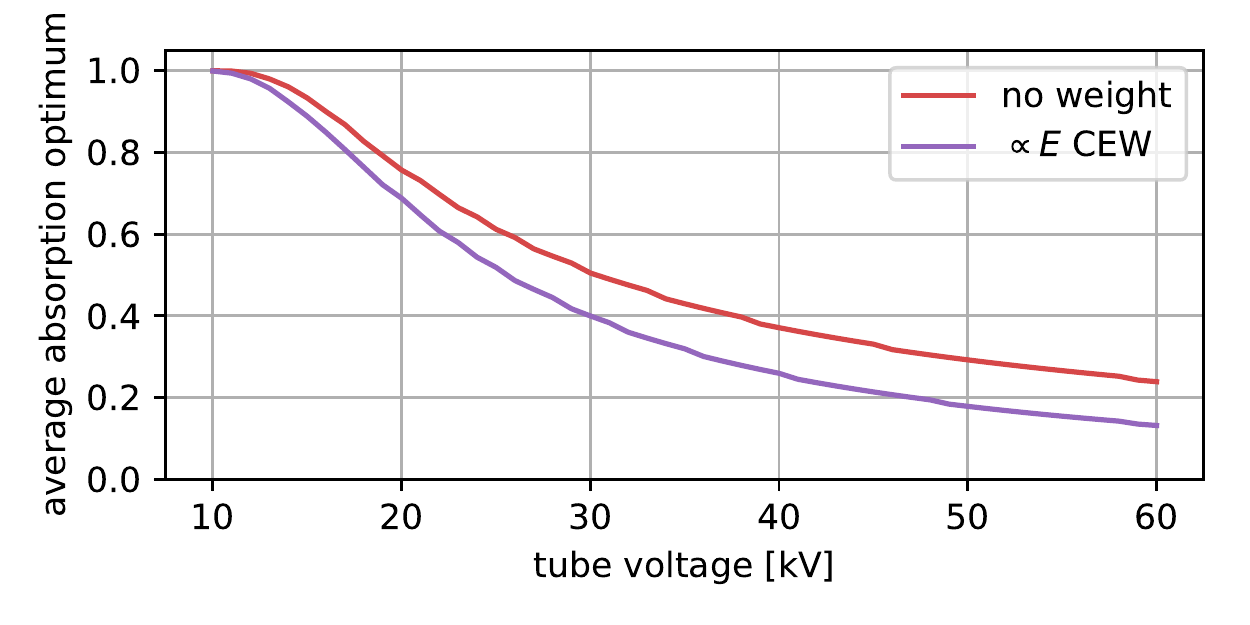}
\caption{Simulated (photo-)absorption $\mathit{SNR}(u_a)$ curves for  different photoabsorber screen thicknesses $t$ (top). 
Corresponding values for the weighted absorption average (upper middle) and energy-dependent absorption efficiency (lower middle). 
The average detector x-ray absorption for the $t$ with the maximal $\mathit{SNR}$ is shown for different tube voltages (bottom).
\label{fig:SNR_screen_thickness_example}}
\end{figure}

It can be seen that weighting down (CEW) or not detecting (DEW) specific photons can increase $\mathit{SNR}$ by large factors. 
Note that the monochromatic $\mathit{SNR}$ is directly proportional to the cumulative intensity of the source spectrum. 

The following simple rules for energy weighting can be seen in Fig.~\ref{fig:SNR_sim_example}: 
For optimal CEW, every additional photon increases $\mathit{SNR}$. 
For optimal DEW no additional photon decreases $\mathit{SNR}$. 
For all other weightings, additional photons can decrease or increase $\mathit{SNR}$. 

If we use this example for a $\mathit{SNR}$ optimization, we can see that very low tube voltages would be optimal if CEW or DEW cannot be implemented. 

We could use the optimal CEW case as $\mathit{SNR}_\text{ideal}$ in eq.~(\ref{eq:DQE}) and the $\propto E$ CEW as $\mathit{SNR}_\text{detected}$ to compute an accurate polychromatic $\mathit{DQE}$ for an indirect detector with ideal absorption and efficient conversion.
It has $\mathit{DQE} \ll 1$ for broad spectra due to its energy weighting.
Computing a polychromatically averaged $\mathit{DQE}$ using eq.~(\ref{eq:def_simple_DQE}) without consideration of the signal strengths would give $\mathit{DQE} = 1$. 
This is where sample-independent $\mathit{DQE}$ models fail.

In real applications, samples usually have varying thicknesses and a setup must be optimized to give high $\mathit{SNR}$ at a combination of thicknesses.

\subsection{Optimal Screen Thickness\label{sec:opt_screen_t}}
$\mathit{SNR}(u_a)$ curves for different thicknesses $t$ of a x-ray absorbing screen (e.g.\ scintillator) are shown in Fig.~\ref{fig:SNR_screen_thickness_example}.
The screen is approximated as an idealized photoabsorber with an absorption constant $\propto E^{-3}$ (no absorption edges). 
All other properties and the sample are the same as above and a source spectrum with a tube voltage of 38\,kV is used in the three upper plots. 
The average absorption values are computed as the weighted detected spectral average of the x-ray absorption efficiency of the detection screen. 
In real cases, the detector MTF depends on the x-ray detection position within the screen and the intensity distribution within the screen is energy-dependent.

We can see that for the "no weight" and the "$\propto E$ CEW" cases, there is an optimal screen thickness. 
Using a thicker screen reduces the $\mathit{SNR}$ due to the lower $\mathit{SDS}$ of the additionally detected higher energy photons. 
The "optimal CEW" and "optimal DEW" methods however prevent this effect. 
In addition, a thinner screen may have a better MTF (this effect is not simulated here). 

In Fig.~\ref{fig:SNR_screen_thickness_example} (bottom) the values of the average x-ray absorption of the detector for the $t$ with the maximal $\mathit{SNR}$ are shown. 
It can be seen that in the absence of optimal CEW or DEW, lower average absorption values are optimal for higher degrees of polychromaticity. 
This is an application of the DEW, as discussed before.

The average absorption for "no weight" is identical with a polychromatically averaged $\mathit{DQE}$ (eq.~(\ref{eq:DQE})). 
Thus the "no weight" and the "$\propto E$ CEW" cases have an optimal $\mathit{SNR}$ at a specific value of the polychromatic $\mathit{DQE}$ which is different from the maximal $\mathit{DQE}$.
Raising the $\mathit{DQE}$ usually has other costs (e.g.\ worse MTF), so that in this case, increasing the $\mathit{DQE}$ beyond its optimum can have direct and indirect disadvantages. 
The "optimal CEW" and "optimal DEW" cases do not benefit significantly if the detector thickness is increased beyond this point.

\section{Discussion\label{sec:ideal_poly_detector}}
\subsection*{Different Use Cases}
The technical capabilities of different x-ray imaging setups vary greatly. 
The samples are also very different: Imaging e.g.\ a \SI{100}{\micro\meter} thick metallic sample at \SI{100}{\nano\meter} resolution or a \SI{500}{\milli\meter} organic sample at \SI{500}{\micro\meter} resolution requires very different setups. 

High resolution x-ray imaging (typical sampling $<$ \SI{10}{\micro\meter}) generally uses lower energies (e.g.\ $<$ 10\,keV) due to the higher x-ray transparency of thinner samples. 
Detecting smaller structures also intrinsically requires a higher $\mathit{SNR}$, so its optimization is generally more important for reducing measurement time than for coarser structures. 
The signal in high resolution x-ray imaging mainly consists of photoabsorption and phase contrast.
In low resolution x-ray imaging, larger samples are investigated.
They require higher x-ray energies.
For the latter, Compton scattering produces a significant part of the absorption signal. 

Within the range of x-ray imaging applications, medical imaging represents a very narrow range. 
Models for image quality which may be appropriate for these cases can be inappropriate in a different context, e.g.\ in material science. 
Imaging scientists may therefore need to use models different from those used in medical imaging (e.g.\ DQE).

$\mathit{SNR}$ spectra can be used to optimize image quality including phase contrast effects \cite{paganin2006,ullherr2018}. 
The latter can be interpreted as a physical highpass filter \cite{paganin.2002}.
For x-ray imaging setups with (sub-)micrometer resolution, phase contrast is often the strongest contribution to the detected signal. 

If one imaging device is used for imaging very different samples, some compromise must be made on the setup components. 
Also, if one sample has very different attenuation lengths along the beam direction, a compromise must be made to optimize image quality for the corresponding different x-ray spectra.

It is not possible to derive general rules that apply to every case. 
For any specific use case, optimizing the $\mathit{SNR}$ spectrum by measurements is required. 
For a set of similar use cases, one can derive general rules. 
Such a set can e.g.\ be imaging with a photoabsorption signal and a bremsstrahlung source. 
A general rule in this example is the fact that only lower x-ray energies contribute to image quality while higher energies deteriorate it.
Additionally, simulations are an important method to discover ways in which an imaging setup may be optimized. 

\subsection*{Structure Size}
Using $\mathit{SNR}$ spectra implies that sample structure (of a specific size) is of interest.  
Because $\mathit{SNR}(u)$ usually strongly decreases to higher $u$ (smaller structures), 
larger structures are always detected well in a specific imaging measurement. 
Smaller structures can only be detected down to a minimal size ("spatial resolution") which is given by the properties of the setup and the measurement configuration. 
For example, the measurement time can be increased to resolve smaller details.

Depending on the structure size of interest, different imaging setups may be optimal. 
This fact is modeled well by $\mathit{SNR}$ spectra. 
Optimal detection depends on what is intended to be detected, which in itself is a decision which needs to be made by the person doing the investigation.

\subsection*{Temporal Measures}
Temporal quantities ($\mathit{SNR}_t$ or $\mathit{DQE}_t$) are defined as properties within a single pixel. 
Due to the fact that the signal in imaging almost always spans multiple pixels, temporal quantities have no reliable quantitative relation to image quality.
Additionally, the value of a temporal quantity depends on the arbitrary choice of the area $A$ of one pixel (or volume and voxel in 3D):
\begin{equation}
\mathit{SNR}_t \propto \sqrt{A}
\end{equation}

Detectors with different pixel sizes thus have intrinsically different $\mathit{SNR}_t$, even if the image quality is identical.\footnote{This can of course be easily corrected by normalizing temporal $\mathit{SNR}$ to pixel area, which is seldom done.}
This is caused by the fact that the incident intensity per detector area is independent of the choice of the pixel size and that temporal $\mathit{SNR}$ is proportional to the square root of the intensity.
$A$ is the area that corresponds to the sampling of the object, not the sampling of the detector.
Changing the x-ray magnification $M$ thus distorts the temporal $\mathit{SNR}$, which unlike image quality is independent of x-ray magnification:
\begin{equation}
\mathit{SNR}_t = \text{const}(M)
\end{equation}

The actual dependence of the image quality is described by:
\begin{equation}
\mathit{SNR}(u) \propto M^2
\end{equation}

which can be confirmed by measuring $\mathit{SNR}(u)$. This is due to the fact that image quality depends on the  detected intensity per object area (not per detector area), which increases with higher $M$.
The smaller area of a object pixel ($A \propto M^{-2}$) at higher magnifications and the constant intensity per pixel leads to an increase of the intensity density.
 
For indirect detectors, temporal measures also neglect the noise correlations that exist (see section~\ref{sec:indirect_detection}). 

The measurement error in a single pixel does not solely determine image quality.
Instead, image quality depends on the image and what we want to detect within.
The use of temporal image quality measures in an imaging context must therefore be considered very carefully, because it can easily result in errors.

\subsection*{Optimizing Imaging Setups}
Designing a good imaging setup or operating it well is conceptually different in a polychromatic use case than in a monochromatic use case. 
For monochromatic imaging, an optimal setup generates the highest possible intensity and detects all photons. 
In the polychromatic case one can apply a CEW and detecting or generating less photons may increase the $\mathit{SNR}$.

If optimal CEW is not possible, then the optimal polychromatic setup needs to detect as many photons as possible from some photon classes and no photons from the other photon classes, see eq.~(\ref{eq:SNR_gain_condition}). 
Depending on the differentiating criterion, this may be very difficult to achieve. 
While a detector can be designed to less efficiently detect x-ray photons of a specific energy range, 
this is much more difficult for x-ray photons that were absorbed at a position in the screen where the MTF is worse (e.g. higher distance from the focus plane).

Counting detectors which can set an upper limit to the x-ray energy of the counted detection events are ideally suited for DEW. 
Setting the upper limit in such a way that photons with comparatively low $\mathit{SDS}$ are not counted may potentially result in a large increase in $\mathit{SNR}$ (see Fig.~\ref{fig:SNR_sim_example}). 


For an indirect detector, the $\mathit{SNR}$ can be increased by choosing a thinner screen for lower energy imaging -- 
even if the absorption efficiency at the relevant energy range may then be slightly lower (see Fig.~\ref{fig:SNR_screen_thickness_example}). 
This effect is separate from the better MTF of the thinner screen.
In many imaging setups, the screen is already thin enough for its use cases due to technical limitations or MTF considerations.

\subsection*{Testable Predictions}
An image quality optimization based on $\mathit{SNR}$ spectra or temporal $\mathit{SNR}$ makes predictions which can be tested directly. 
This is done with a direct measurement, see \ref{sec:measure_SNR} or \cite{ullherr2018} for a method. 
A $\mathit{SNR}$ optimization should ideally be based on measurements for the specific use case. 
Using simulations instead requires that the simulation method was thoroughly tested with direct measurements on the same or a sufficiently similar device. 
This makes it possible to notice errors in the theory or in the assumptions made, which is a basic requirement if one wants to rely on such a model \cite{popper2005}. 

Concerning $\mathit{DQE}$, note that optimizing $\mathit{SNR}$ may lead to a different optimal setting than optimizing polychromatically averaged $\mathit{DQE}$. 
This difference stems from the fact that in practice $\mathit{DQE}$ measurements must use assumptions that may be wrong.

A model can both be tested by physical measurements and also by checking if it is without internal contradictions. 
Here, internal contradictions mainly mean that the model does not actually describe what it is supposed to describe. 
This is for example the case when temporal $\mathit{SNR}$ is used in imaging.

\section{Conclusions}
The concept of signal detection allows a systematic investigation of physical effects that result from varying signal or detection properties for the different detected photons.
Image quality measures like the $\mathit{DQE}(u)$ falsely assume these differences to be unimportant and are therefore unable to model cases where they are. 

Temporal measures like the $\mathit{SNR}_t$ should principally not be used in imaging due to the fact that an image consists of more than one pixel. 

We use $\mathit{SNR}$ spectra as a quantitative model for image quality within the framework of signal detection. 
This allows a robust optimization of image quality in polychromatic x-ray imaging. 
The predictions of this model can and should be tested with direct measurements. 
An image quality optimization on an existing imaging device can be done with direct measurements without the need to correctly model its physical properties.

The main disadvantage $\mathit{SNR}$ spectra have is that they are not an absolute quantitative measure. 
The actual value depends on the object. 
While this is necessary for an accurate model, a quantity that depends only on the imaging setup is more convenient. 
This is solved by simplifying the $\mathit{SNR}$ spectra to a quantity like the detection effectiveness $\mathit{DE}(u)$, eq.~(\ref{eq:DE}), at the cost of a more complex measurement process. 
The $\mathit{DE}(u)$ is not as convenient as an absolutely normalized quantity like the $\mathit{DQE}(u)$, but more robust and accurate.

For polychromatic imaging, we have shown that in a case with varying signal strengths, 
reducing the detected intensity or weight of photons with a relatively low signal detection strength ($\approx$ $\mathit{SDS}$, 
eq.~(\ref{eq:SDS})) can increase the $\mathit{SNR}$ and therefore the image quality. 
While the absolute signal strength does not influence an image quality optimization, relative strength does.
Of course, increasing the detected intensity of photons with a high $\mathit{SDS}$ does increases image quality.
In principle, this result can be applied to any imaging setup.

The model developed here is especially useful for high resolution x-ray imaging and phase contrast imaging. 
We found that depending on the application, direct detectors can have an intrinsically higher $\mathit{SNR}$ than indirect detectors, due to the different energy weightings. 

It is likely that much work analyzing image quality using the $\mathit{DQE}$ or temporal $\mathit{SNR}$
will transfer to the concept of signal detection without large differences in the conclusion. 
In these cases, the benefit will mainly be that the underlying assumptions become clear, 
which in turn may prevent false generalizations. 
In other cases, using the signal detection model is required to accurately model and optimize the image quality of a polychromatic x-ray imaging device.

\subsection*{Acknowledgments}
We would like to thank Theobald Fuchs for test reading and feedback and  Alison Haydock for language help.

\subsection*{Supplementary Material}
\begin{itemize}[topsep=4pt,itemsep=3pt,partopsep=3pt, parsep=3pt]
\item The three Jupyter notebooks (Python) that were used to generate figures, including the simulations.
\item Python code for a $\mathit{SNR}$ spectra evaluation from measurements as described in the \ref{sec:measure_SNR}. Includes examples in two Jupyter notebooks.
\end{itemize}


\bibliography{../../Literatur/literature}


\appendix
\section{Measuring SNR Spectra\label{sec:measure_SNR}}
An important aspect of the $\mathit{SNR}$ Spectra is that they can directly be measured. 
The method described in \cite{ullherr2018} can be used for this purpose---also in applications other than phase contrast.
In the following, we will give a short description of how to apply this method. 

The measurement consists of using an appropriate test object of specific material composition and thickness as sample and 
acquiring a series of $K$ images $\{d_k\}_k$ with this sample. 
The images can either be projections or CT reconstructions. 
Individual images must differ only in their noise realizations. 
The individual exposure times can be shorter than those used in practice; the images should be visibly noisy. 
The averaged image is defined as:
\begin{equation}
d_\text{avg}(x) = \frac{1}{K} \sum_{k=1}^K d_k(x)
\end{equation}

If $D(u)$ is the average power spectrum of the $d_k$, $D_\text{avg}(u)$ the power spectrum of $d_\text{avg}(x)$, 
$\tau$ the integration time of the x-ray detector, then the $\mathit{SNR}$ spectrum for this setup can be calculated as:
\begin{equation}
\mathit{SNR}_\tau(u) = \frac{D_\text{avg}(u) -  \tfrac{1}{K} D(u) - (1-\tfrac{1}{K})A(u)}{ D(u) - D_\text{avg}(u)}\label{eq:measure_snr}
\end{equation}

where $A(u)$ is the power spectrum of the known image artifacts (e.g.\ reference image noise).
This formula works for $K \geq 2$ but gives better results for $K >$ 20.
Note that for computing the power spectra from real images with a FFT, using an appropriate window function is necessary to get an accurate result.

If a reference image is used to normalize the intensity of the projection image, the noise power spectrum of a single reference image can be measured similarly as above with:
\begin{equation}
\mathit{N}_{\tau, \text{ref}}(u) = \frac{D_\text{ref}(u) - D_\text{avg,\,ref}(u)}{1 - \tfrac{1}{K_\text{ref}}}\label{eq:measure_nps}
\end{equation}

This is similar to the denominator of eq.~(\ref{eq:measure_snr}) except for the factor which canceled out.
If the average of the ref images is used for the normalization, it gives a contribution to $A(u)$ of the form:
\begin{equation}
A(u) = ... + \frac{\mathit{N}_{\tau, \text{ref}}(u)}{K_\text{ref}}
\end{equation}

Detector imperfections or other image artifacts not modeled by $A(u)$ can falsely increase the calculated signal and $\mathit{SNR}$. 
These signal artifacts usually appear as a lower limit to $\mathit{SNR}$ at the higher spatial frequencies where the actual $\mathit{SNR}$ becomes very small. 
For a good measurement, this lower limit can be as low as 10$^{-3}$, which is not a practical problem. 

$\mathit{SNR}$ spectra measurements from different samples ("test phantoms") can generally not be compared and their absolute values have limited usefulness. 
The reason is that different amounts of sample structure lead to different $\mathit{SNR}$ spectra.
In many cases, comparability is not needed or the results can be stated as a relative difference between different conditions for the same sample. 

Otherwise, the detection effectiveness spectrum defined in eq.~(\ref{eq:DE}) can be used. 
This quantity can be determined for an experimental setup if the $\mathit{SNR}$ spectrum is measured and the object spectrum $P(u)$ is computed for the object used.
For full comparability, the sample thicknesses must be approximately the same for all measurements that need to be compared.
It is possible to choose a test phantom for which $P(u)$ can easily be computed, e.g.\ one consisting of balls.
This implies that the $\mathit{SNR}$ measurement is done for the CT reconstruction image.

Test phantoms may need to be combined with x-ray filters of different thickness to simulate different sample thicknesses. 
The cumulative thickness of all objects in the beam is an integral part of the definition of the test phantom. 
Any test phantom should satisfy the following properties:
\begin{itemize}[topsep=4pt,itemsep=3pt,partopsep=3pt, parsep=3pt]
\item Its power spectrum should not change with small differences ($<$ \SI{1}{\degree} or $<$ cone angle) in the orientation of the sample to the beam. Round shapes are ideal, e.g.\ balls with a diameter of 20-50 voxels.
\item It should represent a type of sample or a category of samples which will realistically be used.
\end{itemize}

Another possibility of obtaining comparable measurements is to define standardized test phantoms. Examples of such definitions are: 
(1) A geometrically defined object of a specific material.
(2) A random object with a defined statistic, e.g.\ a size distribution of balls and a specific container geometry.

Note that robust measurements of the temporal $\mathit{SNR}$ require the same type of measured data but have a slightly different evaluation. 
Estimating temporal $\mathit{SNR}$ from a region with constant gray values is prone to errors. 

\end{document}